\documentclass{article}

\usepackage{arxiv}
\usepackage{amsmath}
\usepackage{hyperref}       
\usepackage{url}            
\usepackage{booktabs}       
\usepackage{amsfonts}       
\usepackage{nicefrac}       
\usepackage{microtype}      
\usepackage{cleveref}       
\usepackage{graphicx}
\usepackage{cite}
\usepackage{doi}
\usepackage{multirow}
\usepackage{caption}
\usepackage{textcomp}
\usepackage{authblk}

\title{FENCE: Flexible Electric Noise reduCtion Endo-shield for the Suppression of Electromagnetic Interference in Low-Field MRI}

\begin{document}

\author[1]{Julia Pfitzer}
\author[1,2]{Martin Uecker}
\author[1]{Hermann Scharfetter\thanks{Correspondence to \href{mailto:hermann.scharfetter@tugraz.at}{hermann.scharfetter@tugraz.at}}}

\affil[1]{Institute of Biomedical Imaging, Graz University of Technology,
Graz, Austria}
\affil[2]{BioTechMed-Graz, Graz, Austria}

\maketitle

\begin{abstract}
    Electromagnetic interference (EMI) is a significant challenge for low-field MRI systems operating without conventional Faraday-shielded rooms. This interference degrades image quality and limits deployment in space-constrained or electromagnetically noisy environments.
    
    Traditional EMI mitigation approaches include external shields, subject grounding via electrodes, or active noise cancellation requiring synchronized receive channels. These methods either limit portability, introduce patient discomfort, or demand advanced hardware.
    
    In this work, we start from the hypothesis that EMI primarily couples capacitively from the body to the RF coil. We investigated two methods of blocking capacitive coupling while preserving inductive MRI signal detection: First, we employed capacitive segmentation of the RF coil and studied its effect on EMI coupling. Second, we present \textit{FENCE} (Flexible Electromagnetic Noise reduCtion Endo-shield), a novel approach blocking capacitive coupling using flexible PCB shields placed inside the RF coil.  FENCE can be retrofitted to existing RF coils without significant mechanical modifications. 
    Finite element (FE) simulations were used to estimate the expected shielding performance and the impact on RF coil losses prior to practical implementation. 
   
    Testing in various scenarios then demonstrated that the combination of FENCE with segmented solenoid coils is effective against both environmental noise sources and controlled EMI. In phantom experiments, FENCE significantly improved imaging performance and reduced EMI levels to near-baseline levels with ~9\% reduction in coil quality factor (Q factor), showing good agreement with the predictions from the FE simulations.
    In-vivo head imaging confirmed these results across diverse electromagnetic environments significantly improving imaging performance while showing an ~18\% decrease in Q factor. 
    
    FENCE provides a simple method for EMI mitigation in low-field MRI, enhancing image quality while maintaining system portability and accessibility. This approach could help to expand the deployment of low-field MRI systems in low-cost point-of-care applications where conventional shielding is impractical.
\end{abstract}



\section{Introduction}

Electromagnetic interference (EMI) is detrimental for the quality of the reconstructed images and therefore needs to be suppressed as effectively as possible.
Common mitigation methods for EMI in low-field MRI such as Faraday shielding around the subject may not be practical in space constrained environments and also limit the portability of the system.
As fully portable low-field MRI systems are an active area of research \cite{Zhao_2024, Cooley_2020} with portable systems now available in clinical research settings \cite{Yuen_2022}, this represents a challenge that should be addressed.
Several papers have published effective EMI mitigation by subtraction of a weighted sum of filtered reference signals picked up by strategically placed antennas \cite{Bian_2024}.
Some of these techniques are based on a dynamic correction model such as EDITER \cite{srinivas_external_2022} while others rely on deep learning \cite{liu_low-cost_2021,zhao_electromagnetic_2024,zhao_robust_2024, Zhao_2024, Lu_2025}. 
These methods based on reference signals require multiple synchronized receive channels, and may therefore not be an option for every low-field MRI implementation which can lack this functionality \cite{Negnevitsky_2023}.

Generally speaking EMI in low-field MRI can be divided into three different coupling mechanisms:

\begin{itemize}
    \item Conducted interference that couples via cables and electronics to the RF receiver \cite{guallartnaval_2025}.
    \item Inductive Interference that couples via the magnetic field to the RF coil \cite{Yang_2022}.
    \item Capacitive Interference that couples via the electric field to the RF coil \cite{Yang_2022}.
\end{itemize}

For the suppression of EMI in low-field applications, it is essential to identify the primary coupling pathways in order to mitigate  interferences effectively. 
In general conducted interference arising from internal electronics is best mitigated using appropriate filtering and shielding of cables and electronics \cite{guallartnaval_2025}.

For inductive and capacitive coupling, the RF coil in an MRI scanner is a very sensitive antenna tuned to the Larmor frequency of the nuclei in the investigated tissues.
As such it can pick up interference signals from electromagnetic sources with nonzero spectral power within the bandwidth of the coil.
Previous work in the very low-field magnetic field range (<100mT), has determined human body coupling as a major coupling pathway for EMI\cite{Qiao_2025}. 
It was shown, that a majority of EMI doesn't couple directly to the RF coil, instead the human body acts as an intermediate coupling pathway between EMI source and RF coil. 

The question now arises, if this human body coupling happens via a capacitive or inductive pathway or a combination of both.
Notably, many papers that describe active noise cancellation techniques employ primarily inductive reference sensors, i.e. coils. 
However, Srinivas et al.  \cite{srinivas_external_2022} explicitly state that an additional reference signal derived from an electrode attached to the patient improves the EMI suppression considerably, and Yang et al. \cite{Yang_2022} include an additional \textit{ring coil} that is placed around the finger of the subject being measured. These observations suggest that part of EMI may be coupled capacitively from the body of the patient to the receive coil.

In fact, Lena et al. \cite{lena_subject_2025} showed that they could effectively suppress EMI by simply grounding the subject lying in the scanner as well as by adding Faraday shields around the subject. The authors state
that the body acts as a monopole antenna for coupling EMI to the RF coil \cite{Li_2017}. We note that inductively coupled magnetic components of EMI are not expected to be suppressed substantially by grounding, because such coupling to the RF coil via the patient could only happen via eddy currents in the body which cannot be suppressed in this way \cite{bronzino_bmtbook}.

Regarding capacitive and inductive coupling via the body, one can divide these into radiated emission and near-field effects.
Radiated emission is a far-field effect. As such, EMI sources would have to be located at a significant distance from the low-field MRI system \cite{antenna_theory_2005}.
In order to have an effect on the MRI system, these EMI sources would then have to act as high-powered radio transmitters in the medium frequency range.
There are only few such transmitters worldwide and some special services (coastguard etc.) in this frequency range \footnote{mwlist.org}.
Additionally, as human body coupling is already established as the major coupling mechanism, the body would have to act as either a short electrical antenna or small magnetic loop antenna which only in a secondary step couples to the RF coil.
While the body acting as an electrical antenna might be able to pick up radiated emission, for inductive interference this is unlikely to be a more efficient magnetic antenna than the actual RF coil itself.

Based on these observations the main hypothesis of this paper is that the major part of EMI is coupled to the RF coil capacitively via the patient body. To test this hypothesis we investigated two methods for suppressing capacitive coupling between the patient body and the RF coil. The first very effective and well-established method relies on coil segmentation via capacitors as described by Mispelter\cite{Mispelter_2006}. While Mispelter doesn't explicitly address the effect on EMI, segmenting the coil distributes the electric field around the coil, likely reducing capacitive coupling between sample and coil.

The second is the use of a electric shield between the body and the RF coil. Such a shield was used by Gadian and Robinson to determine dielectric losses in NMR experiments on electrically conducting samples \cite{Gadian_1979}. Additionally, previous works in solid-state NMR spectroscopy \cite{Wu_2009,Krahn_2008} showed that such a shield effectively prevents capacitive coupling between RF coils and samples. Park et al. investigated this approach through numerical comparison of designs and experimental verification \cite{Park_2010}. While the authors of those articles aimed at reducing dielectric losses and sample heating at high frequencies, the same concept is also effective for the rejection of capacitively coupled EMI. In contrast, the inductively coupled MRI signal is allowed to pass freely, as long as eddy currents in the shield are prevented as far as possible. This shield must thus be be designed carefully in order not to significantly degrade the performance of the RF coil.

This work explores an approach of using a shield which is placed inside the RF coil in order to prevent capacitive coupling of EMI in low-field MRI.

We call this shield FENCE which is short for \textbf{F}lexible \textbf{E}lectric \textbf{N}oise redu\textbf{C}tion \textbf{E}ndo-shield.

\section{Method}\label{Method}

\subsection{Design of the RF shield}

\begin{figure}[h]
    \centering
    \includegraphics[width=0.5\linewidth]{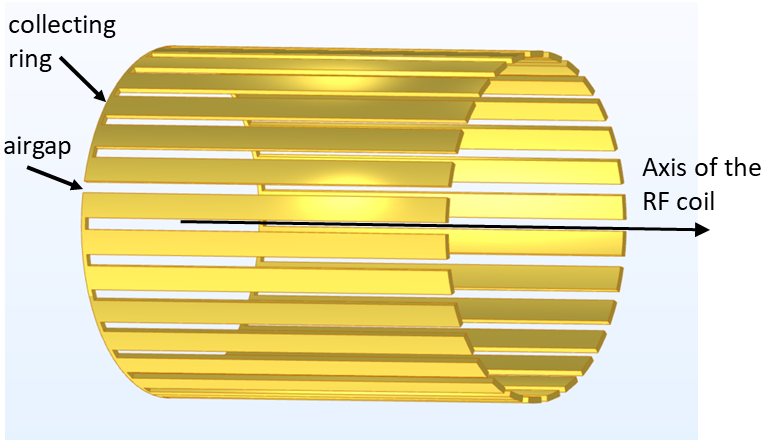}
    \caption{Design of FENCE: The basic structure is a segmented cylindrical shield. The slotted structure prevents eddy current formation. A collecting ring placed on the outside provides continuous electric connection. An airgap in this collecting ring hinders eddy current formation.}
    \label{fig:shieldstruct}
\end{figure}

The shield shall effectively prevent the electric coupling between the coil and the conducting body inside, while not modifying the magnetic $B_{1}$ field. A tube-shaped shield made of a thin layer of highly conductive material would nearly abolish the RF field on the inside at typical low-field MRI Larmor frequencies (approx. 2.1 MHz at 49 mT) even if the layer is considerably thinner than the skin depth. A rough estimation of the damping can be obtained with the approximation formulae given in \cite{conradi_rf_2019} which state that 
\(rt \ll \delta^2\) needs to be fulfilled to achieve low damping of the RF field, \(r\) being the average shield radius, \(t\) the wall thickness and \(\delta\) the skin depth. At usual coil radii, \(t\) needs to be extremely small so as not to perturb the $B_{1}$ field. Thus, at typical layer thicknesses of several tens of micrometers, the shield must be segmented to avoid strong eddy current losses. As the magnetic flux density inside the solenoid RF coil is oriented predominantly along the coil axis, the electric field driving the eddy currents is approximately azimuthal. Therefore the material must be segmented by narrow slots oriented parallel to the axis, as described by Gadian and Robinson \cite {Gadian_1979}. For electric shielding the segments need to be connected to signal ground, thus a collecting ring at one end of the shield is required. To suppress eddy currents in this part the ring should be interrupted by a gap and placed sufficiently far outside the solenoid for low net magnetic flux. Fig. \ref{fig:shieldstruct} illustrates the basic structure of such a shield. 

This is different from strip-shield designs described by Wu et al. \cite{Wu_2009} who explicitly state that the segments are not connected to signal ground. While this design strongly reduces dielectric losses in case of strictly symmetric feeding, it does not shield EMI currents coupled from the environment to the RF coil via the sample.

The quality factor (Q factor) of the RF coil may still be affected by eddy currents flowing tangentially to the shield surface, in particular near the ends of the solenoid, where the magnetic flux density possesses noticeable radial components. These eddy currents grow with the width of the shield segments, thus suggesting to make the strips as narrow as reasonably possible.
There may remain some residual currents which flow between the segments via capacitive coupling, either between adjacent faces or indirectly via the RF coil. However, at 2.1 MHz these contributions are assumed to be small compared to the conductive contributions inside the shielding material.
Electric shielding with a slotted cylinder is, of course, not 100\% efficient because there remains some penetration of the electric field through the slots, thus driving residual EMI currents between the body and the RF coil. In order to further reduce feedthrough to the receiver circuits, we opted for segmenting the RF coil to achieve high EMI suppression by combining both methods. 

\subsection{Finite Element Simulations}

Before designing the shielding hardware, simulations were carried out
by solving Maxwell's equations with the appropriate boundary conditions
with the finite element (FE) method using COMSOL Multiphysics 6.0 (COMSOL, Burlington, MA).
Simulations were performed for two different tasks, i.e.
(1) analysis of the eddy current losses in the shield, and
(2) analysis of the electric shielding efficiency. The eddy current losses were simulated to assess FENCE's impact on coil efficiency. Excessive eddy currents would introduce losses detrimental to coil performance. These losses would degrade the coil's Q factor, which directly affects SNR as, for identical coil geometry, SNR is approximately proportional to the square root of Q \cite{Webb_2023}.
As FENCE employs a slotted structure, its shielding is inherently not perfect. The shielding efficiency simulations were conducted to quantify how effectively the shield functions despite the slotted structure.


Eddy currents were calculated with COMSOL's module 'magnetic field', employing the frequency domain solver with the constitutive equations

\begin{gather}
    \nabla \times \mathbf{H} = J \\
    \mathbf{B} = \nabla \times \mathbf{A} \\
    \mathbf{J} = \sigma \mathbf{E} + j \omega \mathbf{D} + \mathbf{J}_e \\
    \mathbf{E} = -j \omega \mathbf{A},
\end{gather}

with \textbf{H} magnetic field strength, \textbf{J}: current density, $\mathbf{J}_{e}$: injected current density (here zero), \textbf{B}: magnetic flux density, \textbf{A}: magnetic vector potential, \textbf{E}: electric field strength, \textbf{D}: displacement field strength, \(\sigma\): electric conductivity, \(\epsilon\): dielectric permittivity, and \(\omega\): radian frequency. \textbf{H} and \textbf{B} as well as \textbf{E} and \textbf{D} are related through the usual material equations with \(\mu = \mu_{0}\) everywhere and \(\epsilon = \epsilon_{0}*\epsilon_{r}\). \(\epsilon_{r}\) was set to 1 everywhere except in the body cylinder where it was chosen as 80 for water at low frequencies, and in the shield carrier material (values see description of the shield).

\begin{figure}[h]
    \centering
    \includegraphics[width=0.5\linewidth]{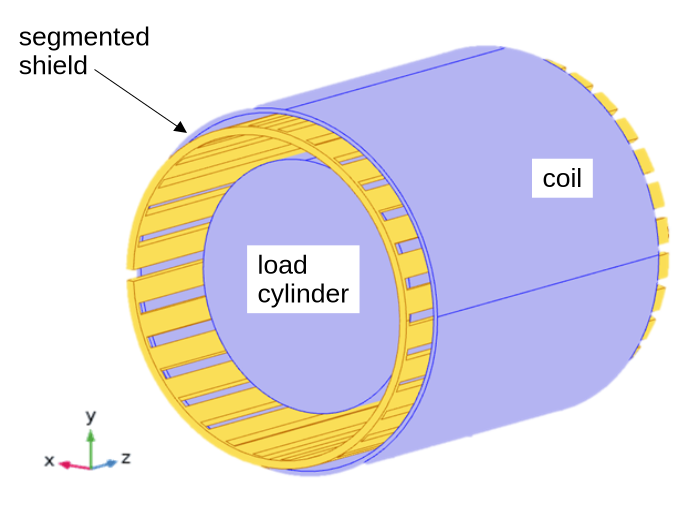}
    \caption{Basic geometry of the magnetic FE model. The segmented shield is placed inside the RF coil. A conductive cylinder simulates coil loading. Simulation parameters are provided in Table \ref{tab:parameters}.}
    \label{fig:modelessentials}
\end{figure}

\begin{table}[ht]
\centering
\caption{Parameters of the FE model}
\label{tab:parameters}
\begin{tabular}{lll}
\hline
\textbf{Symbol} & \textbf{Value} & \textbf{Meaning} \\
\hline
{$R_{c}$} & 155 mm & coil radius \\
{$L_{c}$} & 145 mm & coil length \\
N & 20 & number of turns \\
d & 2 mm & wire diameter \\
t & 2.5 mm & shield thickness \\
{$R_{s}$} & 145 mm & shield inner radius \\
{$L_{s}$} & 180 mm & shield length \\
{$\sigma_{s}$} & \(6 \times 10^7\)S/m & conductivity of the shield\\
{$\sigma_{c}$} & \(6 \times 10^7\)S/m & coil conductivity without skin effect\\
{$\sigma_{body}$} & \ 1.1 S/m & conductivity of the body cylinder\\
{$M$} & 30 & number of slots in the shield \\
{$B$} & 5 mm & slot width \\
{$d_{ring}$} & 2 mm & width of connecting ring\\
t & 18 $\mu$m & thickness of the shield layer\\
f & 2.11 MHz & frequency \\
\hline
\end{tabular}
\end{table}

Only the three essential parts were considered, i.e. the coil, the shield, and a homogeneous loading cylinder for simulating a conductive body (see Fig. \ref{fig:modelessentials}). 
The geometry was defined so as to mimic the experimental setup (Table \ref{tab:parameters}).
The coil was simulated as 'homogenized multiturn coil' with N = 20 turns which is an approximation of the true solenoid structure. As this mode does not explicitly consider skin and proximity effects, the effective wire conductivity was set to the product of the DC-conductivity ($6 \times 10^7$ S/m) and an empirical 'calibration' factor. This factor was chosen such that the calculated Q factor of the unloaded and unshielded coil approached the measured one as closely as possible. As the shield consists of a thin foil of copper, it was modeled as transition boundary condition of a cylindrical shell with radius \(R_s\), axial length \(L_s\) and thickness \(t\) in order to avoid superfine meshing. The respective equations used by COMSOL Multiphysics are
\begin{gather}
    \mathbf{n} \times \mathbf{H}_1 = \mathbf{J}_{s1} \\
    \mathbf{n} \times \mathbf{H}_2 = \mathbf{J}_{s2} \\
    \mathbf{J}_{s1} = \frac{Z_S\mathbf{E}_{t1} - Z_T\mathbf{E}_{t2}}{Z_S^2-Z_T^2}\\
    \mathbf{J}_{s2} = \frac{Z_S\mathbf{E}_{t2} - Z_T\mathbf{E}_{t1}}{Z_S^2-Z_T^2}\\
    Z_S = \frac{-j\omega\mu}{k}\frac{1}{\tan(kt)} \\
    Z_T = \frac{-j\omega\mu}{k}\frac{1}{\sin(kt)} \\
    k = \omega\sqrt{\epsilon + \frac{\sigma}{j\omega}\mu}
\end{gather}

$\mathbf{E}_{t1}$, $\mathbf{E}_{t2}$ refer to the tangential E-fields and $\mathbf{J}_{s1}$, $\mathbf{J}_{s2}$ denote the surface current densities at the two different sides of the layer, respectively\cite{COMSOL_documentation}. 

As an additional boundary condition a magnetic insulation with \(\mathbf{B} = \mathbf{n} \times \mathbf{A}\) was set on a symmetrically placed cylinder surrounding the setup with twice the diameter of the coil and 2.3 times the length of the coil.  

The automatic mode was used for mesh generation allowing for adaptive mesh refinement where necessary. Different starting configurations ('normal', 'fine' ...) were tried and results were accepted if further refinement did not yield further changes of the simulated quantities by more than 1\%. The final mesh consisted of 120165 tetrahedral elements. The
stabilized BiConjugate Gradient using a geometric multigrid approach was used to solve
the equations iteratively.

According to 
\begin{align}
Q &= \frac{j\omega L}{R}
\end{align}

the Q factor of the coil was calculated from the coil's inductance and resistance which, in turn, are derived from the coil voltage and current, respectively. These quantities are provided automatically by COMSOL Multiphysics after solving the field equations.  
The \(B_1\) field homogeneity was quantified by the ratio 
\begin{align}
h &= \frac{|B|_{min}}{|B|_{max}},
\end{align}
where
\(|B|_{max}\) and \(|B|_{min}\) denote the maximum and minimum flux density inside a target cylinder with a radius of 62 mm (80\% of the coil radius) and a length of 72,5 mm (50\% of the coil length). \(h\) was calculated with and without shield by setting the shield conductivity to \(6 \times 10^7\) S/m (copper) and \(0\) S/m, respectively.


Electric fields and complex current densities in the coil were calculated with the module 'electric currents', employing the frequency domain solver with the constitutive equations

\begin{gather}
\nabla \cdot \mathbf{J} = 0 \\
\mathbf{J} = \sigma \mathbf{E} + j \omega \mathbf{D}\\
\mathbf{E} = -\nabla{V},
\end{gather}
.

\begin{figure}[h]
    \centering
    \includegraphics[width=0.5\linewidth]{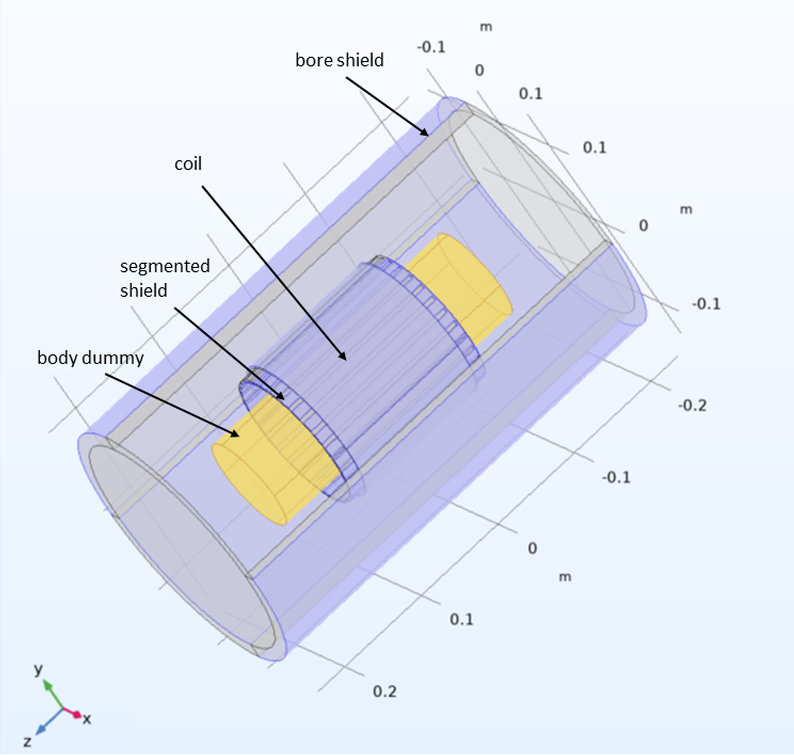}
    \caption{Geometry of the electric FE model for calculating the shielding efficiency.}
    \label{fig:shieldefficiency}
\end{figure}

The geometry of the coil and shield was essentially the same as for the calculation of the eddy currents. 
However, the assembly was enclosed by a grounded cylindrical shell representing the gradient shield which, in the real system, is located between the RF coil and the gradient coils.
Moreover, the body cylinder was replaced by a 'body dummy', i.e. a cylindrical,
highly conducting surface with radius \(R_{b}\) = 50 mm and length \(L_{b}\) = 295 mm, positioned coaxially with all other cylinders and centered symmetrically along the z axis (Fig. \ref{fig:shieldefficiency}). The interior of this cylinder was not meshed to reduce the number of elements. 
The formers of the shield and coil were simulated as hollow cylinders with an effective \(\epsilon_{r}\) of 1.5, corresponding to the used PET material. This value was chosen according to published data which lies in the range of 3 - 4
\cite{Yang2014, Konieczna2010} for pure PET and assuming a filling factor of approximately 50\%. The final mesh consisted of 310973 tetrahedral elements.

To quantify the shielding effect, we defined a shielding factor $\xi$ as
\begin{align}
\xi &= \frac{|I|_{s}}{|I|_{0}},
\label{eq:shieldingfactor}
\end{align}
where \(|I|_{s}\) and \(|I|_{0}\) denote the EMI currents reaching the coil with and without shield, respectively. \(\xi\) can assume values between 0 and 1, 0 representing a completely dense (ideal) and 1 a totally ineffective shield, respectively. Both currents are obtained by integrating \textbf{J} over the coil surface. It should be noted that this setup does not consider the additional effect of RF coil segmentation and thus represents a worst case.

For computing the solution of the current density fields the following boundary conditions (BC) were defined: 
\begin{enumerate}
    \item inner cylinder: injection of unit current ('terminal') 
    \item coil and bore shield: Dirichlet BC with 0 V of potential ('ground')
    \item shield: same as bore when shielding was desired, no explicit BC otherwise
    \item frontal and rear faces of the bore shield and of the inner cylinder: Newman BC with zero current density (’electrical insulation’)
\end{enumerate}

\subsection{RF Coils and Shields}

A specific shield is constructed for each RF coil as the shield is placed on the inside of the RF coil. The section will first cover the types of RF coils used for evaluation 
followed by the different methods for constructing the shields.


For the RF coils, two freely available open-source designs provided by the Open Source Imaging Initiative\footnote{opensourceimaging.org} were chosen.  The first coil (\textit{Phantom Coil}) was intended for phantom imaging \cite{osii_handcoil} whereas the second coil was intended for head imaging (\textit{Head Coil}) \cite{osii_headcoil}. The Phantom Coil lacked capacitive segmentation in its original design which allowed direct comparison between segmented and unsegmented configurations. The Head Coil was used to verify the FENCE approach in vivo after initial tests with the Phantom Coil. Head imaging presents particular EMI challenges, as subject grounding alone is insufficient to obtain usable images \cite{lena_subject_2025}.

\subsubsection{Phantom Coil}
The first coil has 20 turns, a length of \(145,\)mm and a diameter of 155 mm. Litz wire was used to optimize for a high Q factor (Rupalit V155, 5152 x 0.02 mm, 2x63, Pack Litzwire, Germany). 
The coil former was 3D printed using PETG (Extrudr PETG white, FD3D GmbH, Austria) and the litz wire was affixed to the former using UV curing epoxy resin (UV Resin, CHTAWJ, China). 

For EMI mitigation one modification was implemented compared to the original open-source design: The coil was capacitively segmented using three 500 pF capacitors with the value
calculated using Eq. 47 in Ref. \cite{tao_segmented_2023}.
The inductance of the coil was measured at 47.5 µH before segmentation. 
The tuning and matching capacitances were adjusted to obtain proper impedance matching at the Larmor frequency of ~2.1 MHz, as these parameters changed significantly from the original design due to the segmentation. 
Subsequent measurements were carried out with- and without segmentation of the coil to investigate the effect of coil segmentation on EMI coupling.

\subsubsection{Head Coil}
The second coil is an elliptical solenoid with variable pitch designed for head imaging. It consists of 14 turns, measures 125 mm in length with widths of 213 mm and 24 mm, and has
one capacitive segmentation point creating two segments. The same litz wire was used as for the phantom coil. No modifications to the tuning and matching network were required for this coil. It should be noted that in contrast to the phantom coil the head coil uses a symmetric tuning and matching network.

\begin{figure}[h]
    \centering
    \includegraphics[width=0.8\linewidth]{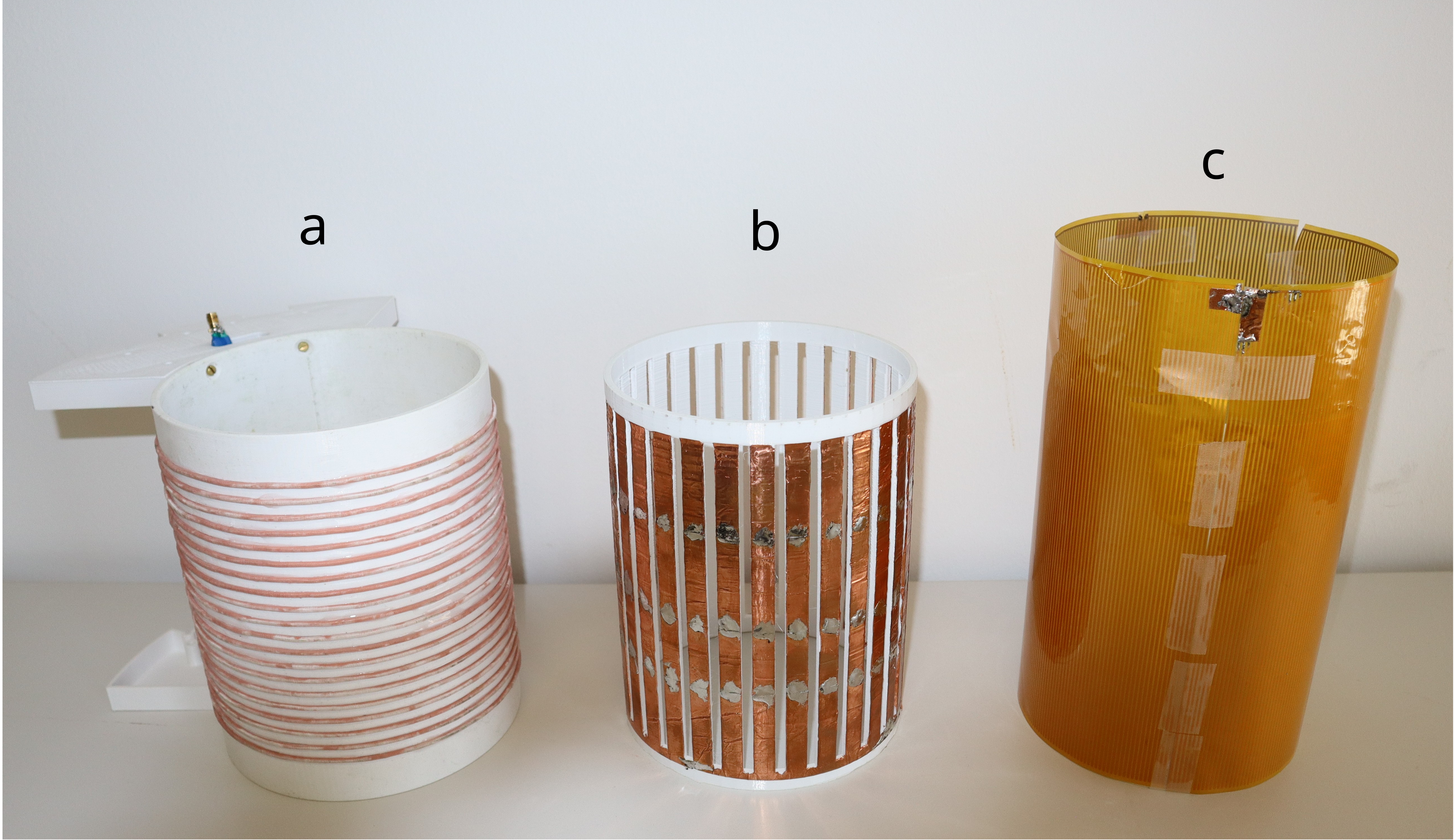}
    \caption{Picture of the phantom coil (a) along with the different shielding implementations: 3D Printed Copper-foil Shield (b) and FlexPCB Shield (FENCE) (c)}
    \label{fig:phantom_coil_setup}
\end{figure}

The RF shields were fabricated using two different manufacturing techniques. One construction method used copper-foil for the shield geometry while the other one used flexible PCB (FlexPCBs). Both shield types were produced for the phantom coil, while only the FlexPCB shield was fabricated for the head coil.
The copper-foil shield was used for initial experiments and for validation of the FE simulations. The FlexPCB approach was developed subsequently as it enables easy construction of thin conductive structures that minimize eddy current formation.
A picture of the phantom coil and the two shields can be seen in Figure \ref{fig:phantom_coil_setup}.

\subsubsection{3D Printed Copper-foil Shield}

The first construction method relies on a 3D-printed PETG former (Extrudr PETG white, FD3D GmbH, Austria) covered with adhesive copper tape (CFT50/20M 1564016 TRU COMPONENTS, Conrad Electronic, Germany) with a copper thickness of $18 \pm 5 \mu\text{m}$. Solder was applied at the overlapping sections of copper tape to improve electrical continuity between sheets. This type of shield was used to verify the FE simulations. This shield will be referred to as \textit{copper-foil shield}.

\subsubsection{FlexPCB Shield (FENCE)}

\begin{figure}[h]
    \centering
    \includegraphics[width=0.6\linewidth]{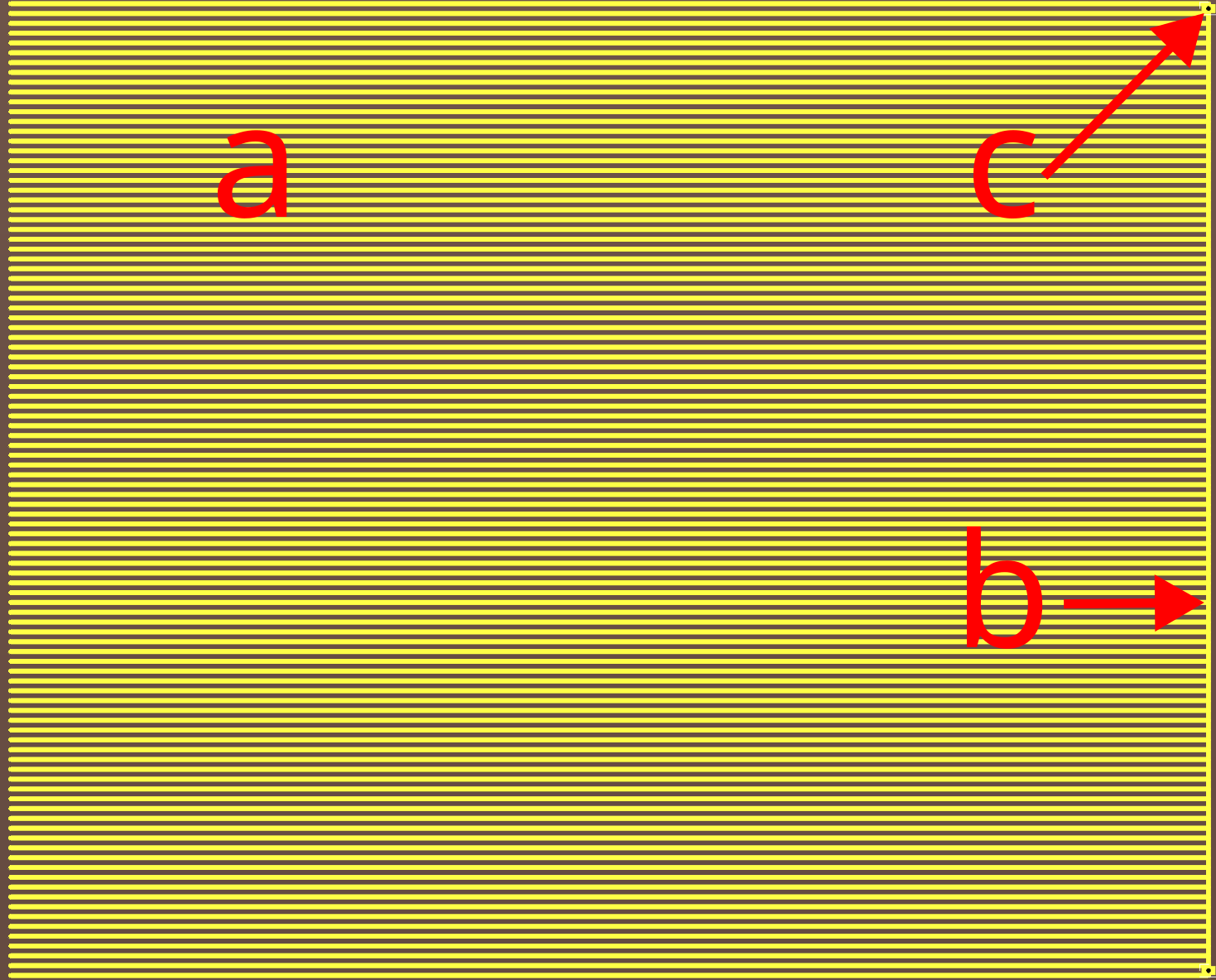}
    \caption{CAD Rendering of one of the FlexPCB segments (FENCE). FENCE can be placed inside RF coils. Each element features 1 mm track width with 1 mm spacing between tracks. The slotted structure  prevents eddy current formation (a). A collector ring positioned on the backside connects the tracks (b).  FENCE elements can be soldered together at their ends to create larger structures (c). The collector should always have a gap, and never form a closed loop to prevent eddy currents in the collector ring.}
    \label{fig:CAD_FENCE}
\end{figure}

The second construction method relies on flexible PCBs that are connected to each other, enabling modular assembly of inner shields. These PCB shields measured 250×200 mm and carried parallel copper strips with 1 mm width and 1 mm spacing. Solder connectors at the edges allowed multiple PCBs to be joined together for modular RF shield construction. The single-layer PCBs were fabricated with 18µm copper thickness on 25µm dielectric substrate (Polyimide) with a total thickness of 0.07mm.
This technique offered significant advantages: the narrower conductive lines reduced eddy current losses, and the flexible PCBs were substantially thinner than 3D-printed formers, preserving more of the imaging volume.
Including the 3D-printed former which ensures the structural integrity of the flexible PCBs, this setup leads to a decrease of about 1.5 mm in usable coil diameter. 
This shield will be referred to as \textit{FENCE}.
The structure of one of the FENCE PCBs is shown in Figure \ref{fig:CAD_FENCE}.
FENCE showed superior mechanical reliability compared to copper foil, which had a tendency to lose electrical connection between overlapping sections when being re-positioned frequently.
The PCBs were externally manufactured at a total cost of 100 EUR for 15 units.
Equipping the phantom coil with FENCE required 3 PCBs whereas the head coil required 4 PCBs.

\newpage

\subsection{Experimental Validation of the FE Simulations} \label{sec:experiment_val}

\begin{figure}[h]
    \centering
    \includegraphics[width=1\linewidth]{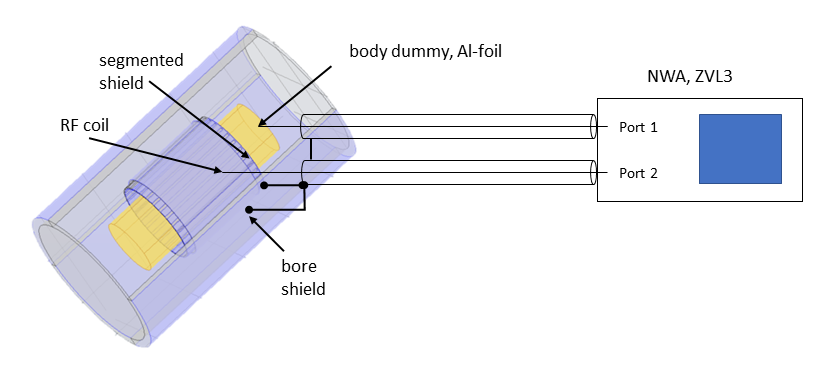}
    \caption{Measurement setup for the determination of the shielding factor.}
    \label{fig:meas_setup_xi}
\end{figure}

For measuring the shielding factor \(\xi\) a hollow polymer cylinder ('body dummy') was wrapped with aluminum foil (Rothilabo 12 \(\mu\), Carl ROTH GmbH, Karlsruhe) and positioned centrally in the RF coil by putting it onto wooden spacers. Both ends of the coil were connected to the core of a coaxial cable (50 \(\Omega)\) with 1 m of length. The cable shield was connected to the bore cylinder and to the slotted shield, using wires as short as possible to minimize parasitic inductances. The other end of the cable was connected to port 2 of a network analyzer (NWA) ZVL3 (Rohde \& Schwarz, Munich, Germany). Port 1 was connected via a cable of the same type to the Al-foil of the inner cylinder, as shown in Fig. \ref{fig:meas_setup_xi}.

The system was calibrated so as to measure the transmission factor \(S_{21}\). In addition the capacitance C between the body dummy and the coil was measured with an LC-meter (BDM-Electronics, Dessau, Germany).
For estimating \(\xi\) the voltage at port 2 was calculated as 
\begin{align}
S_{21} &= \frac{V_{2}}{V_{1}} = \frac{Z_{0}}{2Z_{0}-j\frac{1}{\omega C}}, 
\end{align}
with \(Z_0\) being the characteristic impedance, i.e. 50 \(\Omega\). As the capacitive reactance was much greater than 50 \(\Omega\), one can simplify the expression to
\begin{align}
|S_{21}| \approx Z_{0} \omega C~. 
\end{align}
In this case the NWA can be regarded as an approximate voltage source and the current is then proportional to the \(S_{21}\) value measured. Thus one obtains
\begin{align}
\xi \approx \frac{|S_{21}|_{s}}{|S_{21}|_{0}}, 
\label{eq:xifromS21}
\end{align}
the indices  's' and '0' again denoting the state with and without shield, respectively. 

A phantom was constructed that could be used for loading of the RF coils for Q factor measurements as well as for EMI coupling. 
The dimensions of the phantom matched the cylinder used in the FE simulations. It contained 1.1 l of 0.9\% NaCl solution with 1.5 g/l CuSO$_4$. The conductivity of the solution was measured at 11 mS/cm. Physiological saline solution was used in order to mimic the electrical properties of biological tissue more closely which is relevant in the context dielectric losses and coupling to the RF coil \cite{Gadian_1979}.
EMI coupling to the phantom could be controlled by attaching a thin wire (1 mm diameter, 1 m length). This wire connected to a large copper-foil cylinder (d = 284 mm, h = 440 mm) that captured environmental EMI. A large-surfaced cylinder was used in order to improve EMI pickup and more closely approximate the surface of a human body. An EMI source can be placed within the copper-foil cylinder to intensify interference.
It should be noted that this cylinder only serves as a rough approximation of the human body and therefore does not provide identical EMI coupling properties.

Coils were tuned and matched to the Larmor frequency. No component changes in the tuning and matching circuits were required after inner shield installation, though variable capacitors required adjustment.

The Q factor of each coil was measured using the double coil method described in \cite{Mispelter_2006}. 
All Q factor measurements were conducted inside the low-field MRI scanner bore within a 284 mm diameter copper shield connected to RF ground.
Measurements were taken in both loaded and unloaded configurations, before and after adding inner shields. For the phantom coil, measurements included both the copper-foil shield and FENCE shield variants.

All Q factor measurements for the phantom coil were carried out with capacitive RF coil segmentation. The head coil Q factor was measured in both loaded and unloaded conditions with the flexible PCB shield before conducting in-vivo experiments.

\subsection{MRI Measurements}

To demonstrate the effectiveness of the inner shielding approach combined with coil segmentation, both phantom and in-vivo experiments were conducted on a low-field MRI scanner.


The MRI scanner used for the phantom and in-vivo experiments was built locally based
on the open-source OSI$^2$ ONE design \cite{{osii_one}}. It incorporates a MaRCOS-based console \cite{Negnevitsky_2023} \cite{ocra}
with hardware designs from the Open Source Imaging Initiative \cite{osii_console}.
The MaRCOS-based console features a maximum of two synchronized receive channels. Software-based solutions like EDITER are therefore not possible with this console.

Both gradient and RF amplifiers are based on open-source projects from the same initiative \cite{osii_rfpa} \cite{osii_gpa}. The system utilizes a permanent magnet array \cite{osii_magnet} with a field strength of ~49 mT (2.11 MHz at 22 °C) shimmed to a homogeneity of 1600 ppm over a 200 mm diameter of spherical volume (DSV) as well as in-house built gradient coils (x: $0.438\frac{mT}{m \cdot A}$, y: $0.910\frac{mT}{m\cdot A}$, z: $0.597\frac{mT}{m\cdot A}$).
The active transmit-receive (TXRX) switch followed another open-source design from OCRA \cite{a4im_txrx}. For the low noise amplifier (LNA), a commercial product was used (ABL0100-00-6010, Wenteq Microwave Corporation, United States).
The gain of the LNA was measured at 58.5 dB at the Larmor frequency of the system.
Control of the MRI scanner was performed using MaRGE, an open-source console software offering a wide range of pulse sequences and hardware calibration options \cite{Algar_n_2024}.

One modification to the original OSII$^2$ ONE design was the addition of an RF shield between the gradient coils and RF coil (\textit{gradient shield}). This shield consists of a solid copper cylinder with 0.8 mm thickness and 284 mm diameter, placed within the bore and connected to RF ground. This shield prevents EMI coupling from the gradient coils to the RF coil. Neither the magnet assembly nor the electronics tray were enclosed in a Faraday shield.

Conducted interference was mitigated prior to the MRI experiments by shielding of cables and filtering and comparing measured noise levels to a 50 $\Omega$ baseline.


For the MRI measurements a number of different protocols were carried out to determine the level of EMI and also the effect of the FENCE shields on the image SNR. Before each measurement the coils were tuned and matched to a minimum $S_{11}$ value
of -15 dB.

\subsubsection{Noise Spectrum}
A 50 kHz wide noise spectrum was acquired by executing a Rapid Acquisition with Relaxation Enhancement (RARE) sequence without RF pulses, recording 1000 noise scans. The Fourier transform was computed for each noise scan, and the mean of the resulting spectra was calculated.

\subsubsection{Noise Factor}
A noise factor was measured using the \textit{Noise} protocol in MaRGE, which samples the ADC signal over a specified bandwidth. For these measurements, an acquisition bandwidth of 50 kHz was used.
Additionally, a baseline noise spectrum, where the TXRX switch was terminated with 50 $ \Omega$, was obtained.
As all cables as well as all electronics enclosures were shielded, it is not expected that EMI sources within the scanner environment would significantly couple to the receive chain during the baseline noise measurements.
The root mean square of the real part of the noise signal \textit{Noise Level} was used to compare the different noise levels. 
The \textit{Noise Factor} could then simply be calculated from the ratio of the measured  \textit{Noise Level} and \textit{Baseline Noise}.
\begin{align}
    \textit{Noise Factor} = \frac{\textit{Noise Level}}{\textit{Baseline Noise}}
\end{align}

\subsubsection{Images}

The images were directly reconstructed using an inverse fast Fourier transform.
No corrections for gradient or $B_0$ inhomogeneities were applied.
The high Q factor of the used RF coils means that the bandwidth of the coil can be less than the bandwidth of the image acquisition. Methods to mitigate these effects are described by Webb and O’Reilly \cite{Webb_2023} who obtain a coil sensitivity profile by fitting a polynomial function to a noise scan. The images are then corrected by multiplying with the inverse of this function. However this approach is only valid for situations without major EMI sources. If EMI sources with non-uniform noise spectra are present, this method no longer approximates a sensitivity profile of the RF coil, but rather the spectral distribution of the EMI source. As EMI was actively introduced in the different experiments, the images were therefore not corrected for RF coil bandwidth related shading.

\subsubsection{SNR}
SNR was calculated for the reconstructed images using four noise ROIs placed outside the imaging region (orange) and a central ROI for signal mean estimation (pink). SNR was calculated as the ratio of mean signal ($S$) to mean standard deviation of the noise ROIs ($N$). SNR was calculated as 
\begin{equation}
    SNR = \frac{S}{N \cdot \sqrt{2}}~.
\end{equation}
The noise level was scaled with a factor of $\sqrt{2}$ due to the low SNR which gives rise to a Rician noise distribution \cite{Gudbjartsson_1995}.
The noise ROIs were placed at more central locations in the image to reduce effects of digital filters which create a signal roll-off at the edges of the image.
Digital filter effects in combination with the previously discussed bandwidth related shading introduces a bias in the SNR calculation. However as this study compares different EMI mitigation strategies and scenarios, the relative change in SNR can still be determined accurately.

Phantom measurements provided a repeatable setup allowing comparison between strong and minimal EMI coupling conditions using the same sample. The used phantom is described in Section \ref{sec:experiment_val} and the phantom coil was used as an RF coil.
    
\begin{figure}[h]
    \centering
    \includegraphics[width=\linewidth]{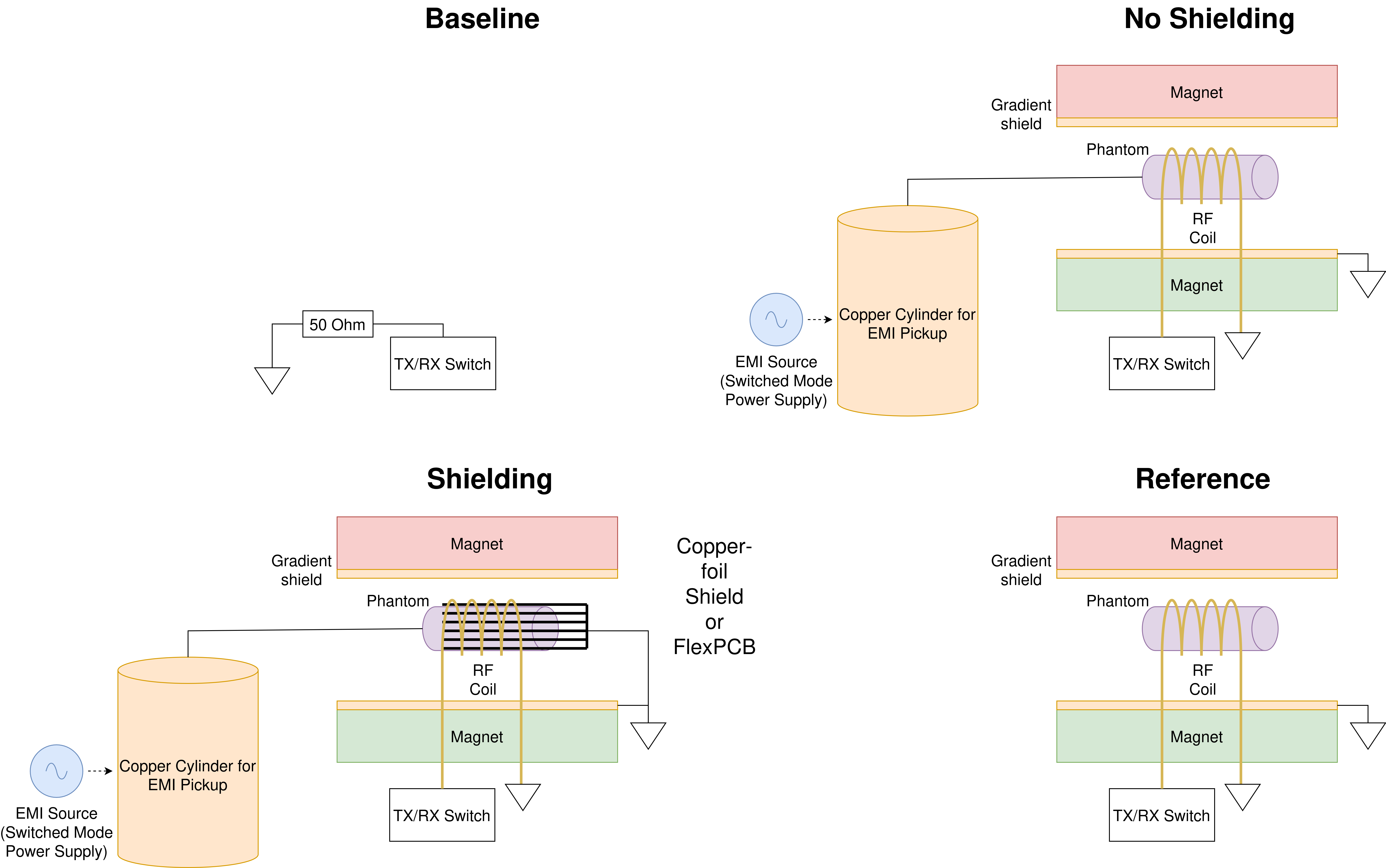}
    \caption{Experimental setup for phantom measurements: \textbf{Baseline:} TXRX switch terminated with 50$ \Omega$. \textbf{No Shielding:} phantom connected to copper cylinder for EMI pickup without shield. This measurement was conducted both with and without capacitive RF coil segmentation. \textbf{Shielding:} inner shields (copper-foil / FENCE) added to RF coil for EMI suppression. \textbf{Reference:} measurement without EMI coupling to phantom and no inner shield.}
    \label{fig:phantom_mrisetup}
\end{figure}

Figure \ref{fig:phantom_mrisetup} shows the different measurement configurations for the phantom measurements.
For the \textit{No Shielding} configuration, the RF coil was tested both with and without capacitive segmentation. This comparison demonstrated that capacitive segmentation alone reduces EMI coupling to the RF coil.
The configurations \textit{Shielding} and \textit{Reference} were only conducted with a segmented RF coil. For these measurements the RF coil was first equipped with the copper-foil shield and then the FENCE shield. 
The reference measurement, where no additional EMI was coupled to the phantom, was conducted to analyze the potential SNR loss due to the presence of the inner shielding structures. It should be noted however that residual EMI could still couple to the phantom, as the measurements were not conducted in a fully Faraday shielded room.

The mean SNR and SNR standard deviation over 10 slices was calculated for each configuration.  

The 3D RARE sequence available in the MaRGE console software was used for imaging with the parameters listed in Table \ref{tab:mri_params_phantom}.

\begin{table}[ht]
\centering
\caption{Phantom MRI Acquisition Parameters}
\label{tab:mri_params_phantom}
\begin{tabular}{ll}
\hline
\textbf{Parameter} & \textbf{Value} \\
\hline
TR/echo spacing/echo train length & 400 ms / 20 ms / 10 \\
Field-of-view & 160 × 160 × 160 mm$^3$ \\
Data points & 80 × 80 × 32 \\
Voxel size & 2 × 2 × 5 mm$^3$ \\
Acquisition bandwidth & 20 kHz \\
RF pulse length & 150 $\mu$s / 300 $\mu$s \\
Trajectory & Cartesian Inside-out \\
Averages & 1 \\
Acquisition Time & 2 minutes \\
\hline
\end{tabular}
\end{table}

Due to eddy currents on the gradient shield, the k-space center shifted in the readout direction. To mitigate this, the shift was calibrated using phantom measurements, and the readout window was adjusted accordingly along with an extension of the readout gradient.
First order shimming was performed using the gradient coils.


In-vivo experiments involved head imaging using the head coil in four different configurations:

\begin{itemize}
    \item No subject grounding
    \item Grounding the subject using a single ECG electrode (Cleartrace-2 Radiotranslucent ECG-Electrode, ConMed, United States) placed on the upper arm
    \item FENCE: Using the FlexPCB inner shield
    \item FENCE and subject grounding.
\end{itemize}
A switched-mode power supply (ADLX65YSLC3A, Lenovo, China) connected to the power grid was held by the subject to introduce additional EMI during the MRI measurement. 

To further evaluate FENCE efficacy, additional head imaging experiments were performed in non-shielded configuration and with FENCE under various other EMI conditions.
Environmental EMI sources were introduced by placing the scanner in close proximity ($< 2 m$) to active equipment: first an operational 3D printer (V-Core 4, RatRig, Portugal), and subsequently an operational large-volume field-mapping robot (COSI Measure \cite{Han_2017}). 
Controlled EMI sources were introduced by positioning a copper-foil cylinder (d = 284 mm, h = 440 mm) one meter from the subject. A signal generator (DG4162, RIGOL Technologies, China) was connected to the cylinder and introduced EMI in two different scenarios, broadband EMI ($\frac{28\mu V}{\sqrt{Hz}}$) and single-frequency EMI (5$ mV_{pp}$ at 2.111 MHz). 
As measurements were conducted in an unshielded lab environment, various unknown noise sources were present throughout the experiment series with undetermined contributions.
A picture of the measurement setup can be found in the supplementary material (S1).

For imaging, the 3D RARE sequence was used again with the parameters listed in Table \ref{tab:mri_params_invivo}.

\begin{table}[ht]
\centering
\caption{In-Vivo MRI Acquisition Parameters}
\label{tab:mri_params_invivo}
\begin{tabular}{ll}
\hline
\textbf{Parameter} & \textbf{Value} \\
\hline
TR/echo spacing/echo train length & 500 ms / 20 ms / 5 \\
Field-of-view & 240 × 200 × 150 mm$^3$ \\
Data points & 120 × 100 × 30 (RO, PH1, PH2)\\
Voxel size & 2 × 2 × 5 mm$^3$ \\
Acquisition bandwidth & 20 kHz \\
RF pulse length & 250 $\mu$s / 500 $\mu$s \\
Trajectory & Cartesian Inside-out \\
Averages & 2 \\
Acquisition Time & 10 minutes \\
\hline
\end{tabular}
\end{table}

In-vivo experiments were carried out with approval from the local ethics committee and written informed consent was obtained from the subject prior to the measurements.



\section{Results}\label{Results}

\subsection{Validation of the FE Simulations}

The simulated and measured values for the phantom coil Q factor in loaded and unloaded configurations  are shown in Table \ref{tab:qfactor}.
The unloaded Q factor for the copper-foil shield dropped significantly by over 40\% compared to the configuration without any shield. For FENCE this decrease is lower at 20\%. When observing the loaded Q factor the copper-foil shield dropped by 25\% while for the FENCE the reduction was 9\%. 

\begin{table}[ht]
\centering
\setlength{\tabcolsep}{4pt}
\caption{Q factor measurements for the segmented RF coil with different shield configurations}
\label{tab:qfactor}
\begin{tabular}{lccccc}
\hline
\multirow{2}{*}{\textbf{\shortstack{Shield\\Configuration}}} & \multicolumn{2}{c}{\textbf{Unloaded Q}} & \multicolumn{2}{c}{\textbf{Loaded Q}} \\
\cline{2-5}
& \textbf{Simulated} & \textbf{Measured} & \textbf{Simulated} & \textbf{Measured} \\
\hline
No Shield & 512 & 509 & 289 & 267 \\
Copper-foil Shield & 300 & 292 & 208 & 199 \\
FENCE & - & 426 & - & 245 \\
\hline
\end{tabular}
\end{table}

The measured values for the head coil with FENCE and with no shielding can be found in Table \ref{tab:qfactor_head}. With FENCE, the unloaded Q factor dropped by 20\% whereas the loaded Q factor decreased by 18\%.

\begin{table}[ht]
\centering
\caption{Q factor measurements of the head coil with different shield configurations}
\label{tab:qfactor_head}
\begin{tabular}{lcc}
\hline
\textbf{Shield Configuration} & \textbf{Unloaded Q} & \textbf{Loaded Q} \\
\hline
No Shield & 428 & 392 \\
FENCE & 340 & 323 \\
\hline
\end{tabular}
\end{table}


In the simulation the factor h which measures the \(B_1\) field inhomogeneity changed from 0.638 to 0.649 when introducing FENCE. As expected, this change is very small and not significant for imaging. 


The capacitance C between coil and body dummy was 16 pF, thus justifying the simplification used for Eq. \ref{eq:xifromS21} at 2 MHz.
The simulated shielding factor \(\xi\) according to \ref{eq:shieldingfactor} was 0.073 with \(\epsilon_{s} = 2\), 0.075 with \(\epsilon_{s} = 1.5\) and 0.075 with \(\epsilon_{s} = 1\), while the measurement yielded a factor of 0.064.

\subsection{MRI Measurements}

\begin{figure*}[!t]
    \centering
    \includegraphics[width=1.0\linewidth]{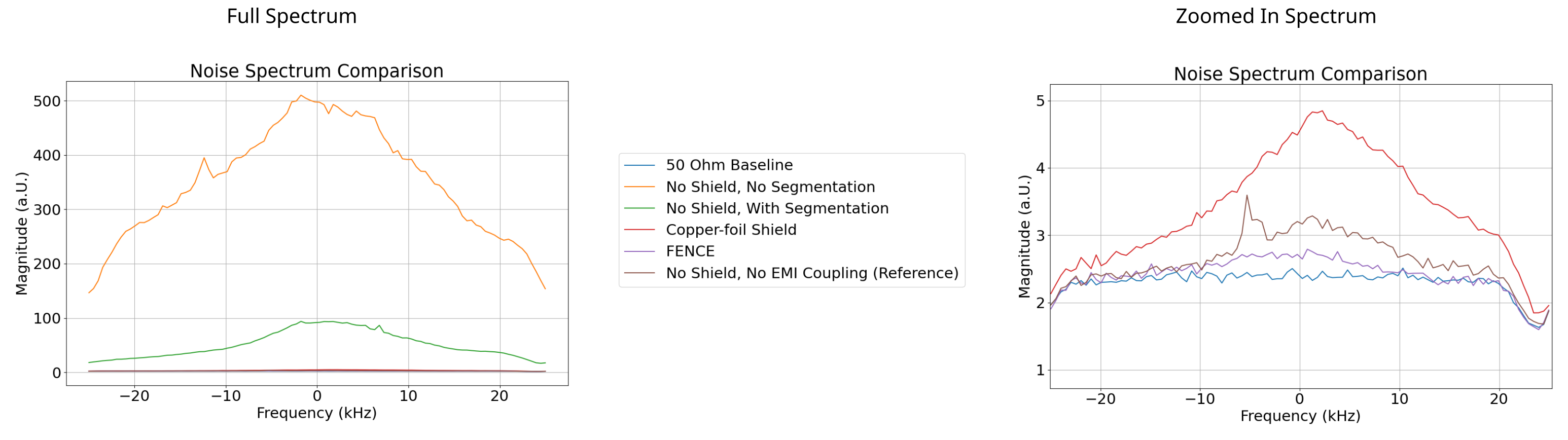}
    \caption{Noise spectrum measurements for different EMI mitigation configurations.}
    \label{fig:noise_spectrum}
    
    \vspace{1em}
    
    \centering
    \begin{tabular}{lcc}
    \hline
    \textbf{EMI Mitigation Configuration} & \textbf{Noise Level (µV)} & \textbf{Noise Factor} \\
    \hline
    No Shield, No Segmentation & 28341.8 & 154.45 \\
    No Shield, With Segmentation & 4486.6 & 24.45 \\
    Copper-foil Shield & 275.9 & 1.5 \\
    FENCE & 192.1 & 1.05 \\
    No Shield, No EMI Coupling (Reference) & 211.1 & 1.15 \\
    \hline
    \end{tabular}
    \captionof{table}{Measured \textit{Noise Level} and \textit{Noise Factor} for different EMI mitigation configurations. The \textit{Noise Factor} is scaled to a 50 $\Omega$ \textit{Baseline Noise} which was $183.5\,\mu V$.}
    \label{tab:noise_levels}
    
\end{figure*}

In the phantom, the noise measurements showed a six-fold reduction of the noise level by RF coil segmentation and another reduction by a factor of 16 and 23 for the copper-foil shield and FENCE, respectively (Figure \ref{fig:noise_spectrum} and Table \ref{tab:noise_levels}).
In the noise spectra for the 50 $\Omega$ baseline an uneven noise distribution at the edges can be observed due to digital filter effects.

\begin{figure*}[!t]
    \centering
    \includegraphics[width=\linewidth]{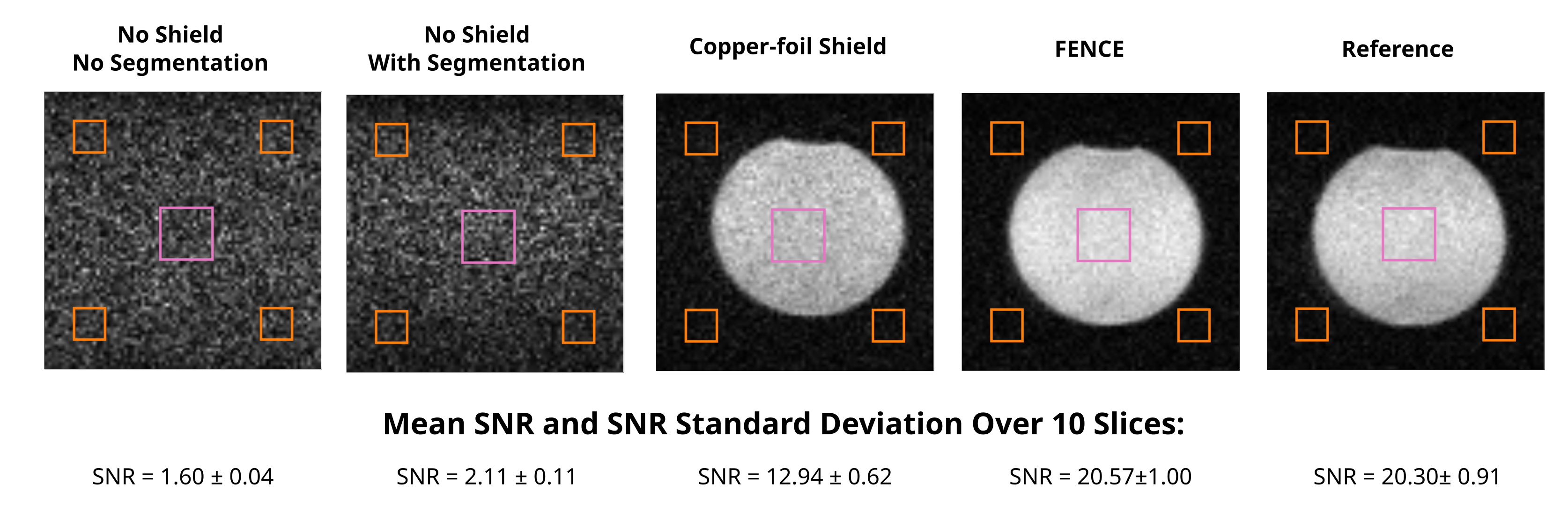}
    \caption{Single slice of an MRI phantom measured with 3D RARE under different shield configurations. SNR was calculated using four noise ROIs placed outside the imaging region (orange) and a central ROI for signal mean estimation (pink). SNR was calculated as the ratio of mean signal to scaled standard deviation of the noise ROIs. Mean SNR and SNR standard deviation were calculated across 10 slices for each configuration. The addition of inner shields (Copper-foil and FENCE) significantly improved SNR. The reference image on the right was acquired without additional EMI coupling to the phantom and no inner shields. }
    \label{fig:Phantom_SNRplot}
\end{figure*}

Figure \ref{fig:Phantom_SNRplot} shows the SNR for the phantom measurement.
SNR measurements without shielding yielded images indistinguishable from  noise, with the noise being substantially reduced when using a shield. The image acquired with FENCE comes close to the reference image acquired without EMI coupling.

\begin{figure*}[!t]
    \centering
    \includegraphics[width=1.0\linewidth]{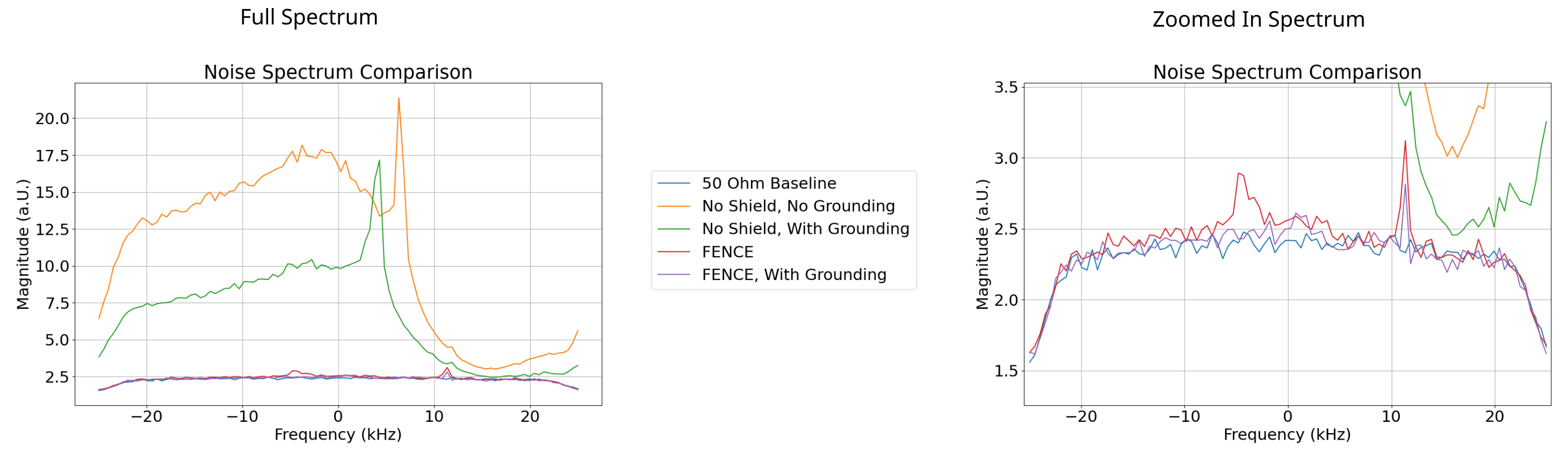}
    \caption{Noise spectrum measurements for different EMI mitigation configuration.}
    \label{fig:in_vivo_noise_spectrum}
    
    \vspace{1em}
    
    \centering
    \begin{tabular}{lcc}
    \hline
    \textbf{EMI Mitigation Configuration} & \textbf{Noise Level (µV)} & \textbf{Noise Factor} \\
    \hline
    No Shield, No Grounding & 1150.0 & 6.2 \\
    No Shield, With Grounding & 749.0 & 4.04 \\
    FENCE, No Grounding & 194.3 & 1.05\\
    FENCE, With Grounding & 188.8 & 1.02 \\
    \hline
    \end{tabular}
    \captionof{table}{Measured \textit{Noise Level} and \textit{Noise Factor} for different EMI mitigation configurations. The \textit{Noise Factor} is scaled to a 50 $\Omega$ \textit{Baseline Noise} which was $185.6\,\mu V$.}
    \label{tab:in_vivo_noise_levels}
    
\end{figure*}

In-vivo, the noise measurements showed a significant EMI mitigation by the inner shields (Figure \ref{fig:in_vivo_noise_spectrum} and Table \ref{tab:in_vivo_noise_levels}).    
When not using FENCE, grounding leads to a substantial reduction of the noise level.
FENCE without grounding leads to a reduction by a factor of six which is slightly improved with additional grounding.

\begin{figure*} [!t]
    \centering
    \includegraphics[width=\linewidth]{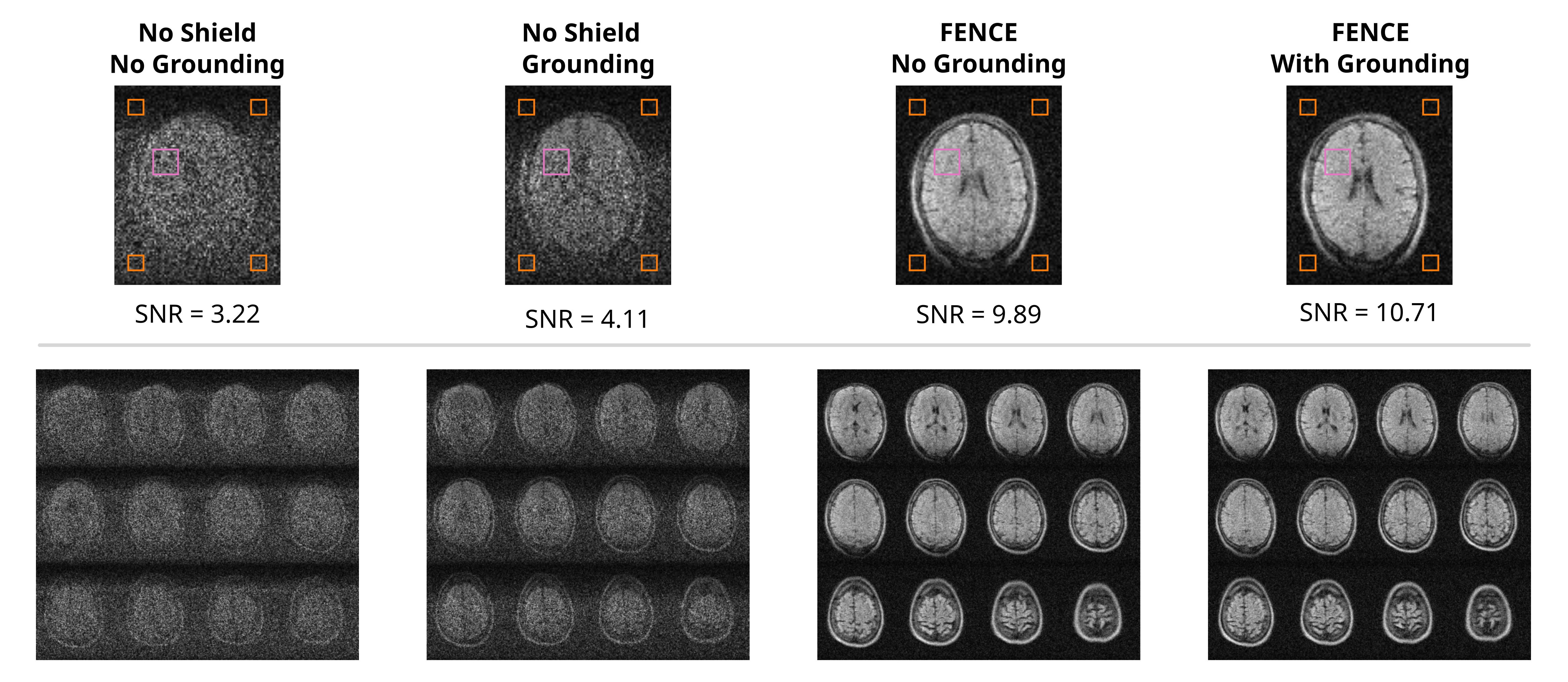}
    \caption{3D RARE images of a volunteer head. SNR was calculated using four noise ROIs placed outside the imaging region (orange) and a central ROI for signal mean estimation (pink). SNR was calculated as the ratio of mean signal to scaled standard deviation of the noise ROIs. Twelve slices are displayed for the different configurations. }
    \label{fig:in_vivo_SNR}
\end{figure*}

Figure \ref{fig:in_vivo_SNR} shows the SNR for the different in-vivo measurements. 
Again, grounding has some effect when not using a shield, while FENCE alone  substantially increases SNR. Additional grounding yields only minimal improvement.

\begin{figure*}[!t]
    \centering
    \includegraphics[width=1.0\linewidth]{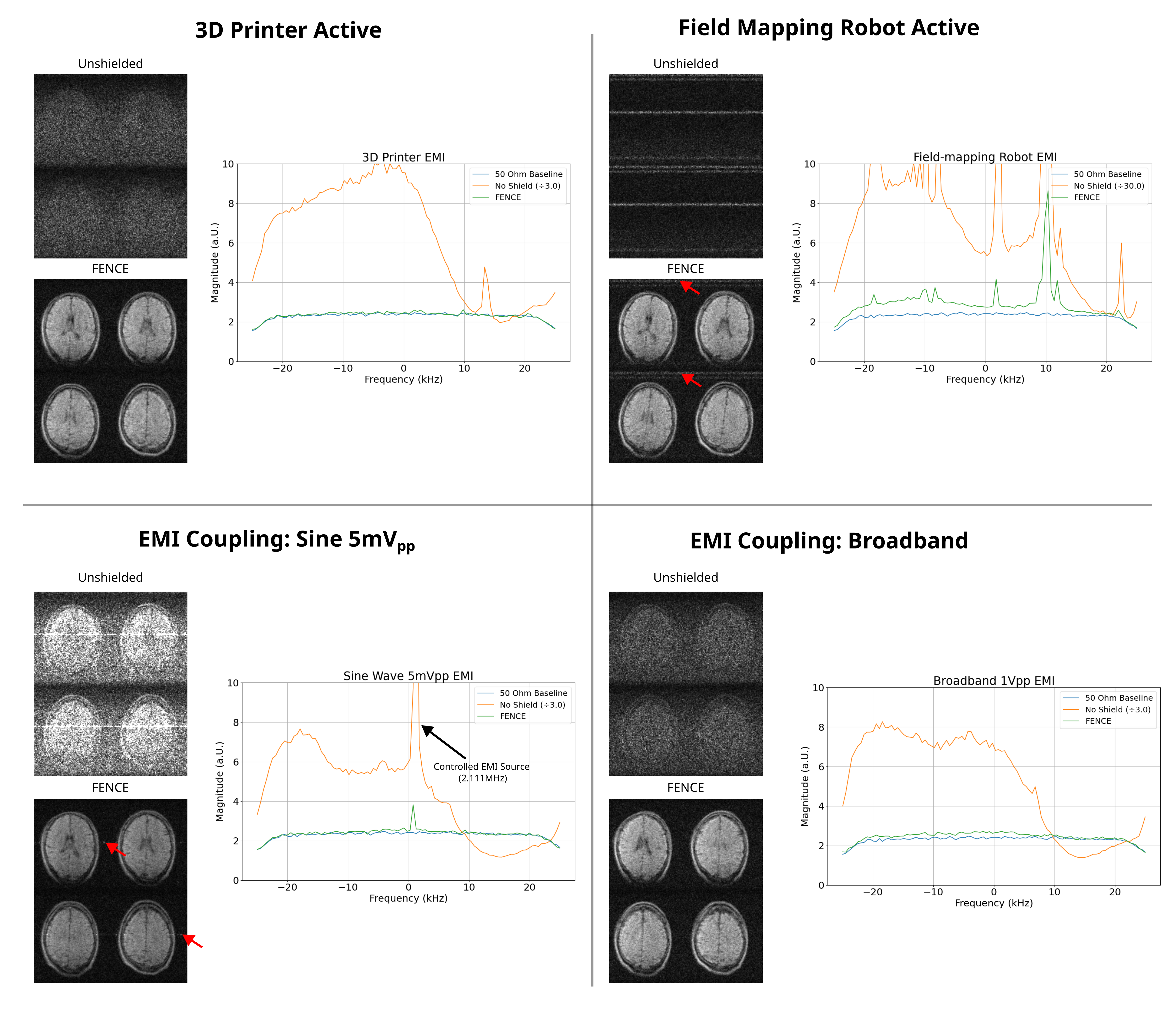}
    \caption{Comparison of FENCE effectiveness across different EMI scenarios. Red arrows highlight subtle image artifacts. Noise spectra were scaled as indicated in the legend to accommodate varying noise levels within a single display range. The maximum unshielded value reached 1080  for the \textit{Field-mapping Robot EMI} and 70 for the \textit{Sine Wave 5mVpp EMI}.}
    \label{fig:robustness_comparison}
\end{figure*}

Figure \ref{fig:robustness_comparison} and Table \ref{tab:emi_comparison} show the effect of the different EMI sources, demonstrating effective suppression of EMI when using FENCE with only minor residual artifacts visible for the field-mapping robot and the sine wave generator.

When operating the field-mapping robot in close proximity to the RF coil, single-frequency interference peaks reached levels exceeding 400 times baseline (Figure \ref{fig:robustness_comparison}) and 100 times broadband noise factor values (Table \ref{tab:emi_comparison}), creating visible image artifacts despite the FENCE shield (Figure \ref{fig:robustness_comparison}).

\clearpage

\clearpage

\begin{table*}[!ht]
\centering
\caption{Performance comparison of EMI mitigation configurations across various interference sources. The \textit{Noise Factor} is scaled to a 50$\Omega$ \textit{Baseline Noise} which was $185.63\mu V$.}
\label{tab:emi_comparison}
\begin{tabular}{lcccc}
\hline
\textbf{EMI Source} & \textbf{Shielding Type} & \textbf{Noise Level (µV)} & \textbf{Noise Factor} & \textbf{SNR} \\
\hline
\multirow{2}{*}{3D Printer Active} & No Shield & 1956.2 & 10.54 & 2.02 \\
 & FENCE & 187.4 & 1.01 & 10.55 \\
\hline
\multirow{2}{*}{Field Mapping Robot Active} & No Shield & 20832.9 & 112.23 & 1.48 \\
 & FENCE & 240.1 & 1.29 & 7.00 \\
\hline
\multirow{2}{*}{EMI Coupling: Sine (5mVpp)} & No Shield & 1262.6 & 6.8 & 2.96 \\
 & FENCE & 195.19 & 1.05 & 11.41 \\
\hline
\multirow{2}{*}{EMI Coupling: Broadband} & No Shield & 1432.9 & 7.719 & 2.36 \\
 & FENCE & 193.9 & 1.04 & 10.23 \\
\end{tabular}
\end{table*}

\clearpage

\section{Discussion}

We have shown that electrical shielding together with the use of segmented RF coils considerably improves EMI immunity of a low-field MRI under normal lab conditions  even without the use of any active noise suppression.  This confirms our hypothesis of capactive coupling of EMI from the body of the patient to the RF coil.

The use of segmented coils alone already strongly reduces EMI susceptibility, which is expected as the external electric field of such a coil possesses high symmetry with many sign changes along the main coupling direction with the internal conducting body. This then reduces coupling substantially, which is well known already in the context of RF coil losses.
We then showed that the use of an electric shield further suppresses EMI to an extent which makes
in-vivo imaging possible without active noise suppression.

For the design of the shield, there is a trade-off
between shielding efficiency and the degradation of the coil's Q factor due to eddy currents.  The
combined use with a segmented coil allows the use
of a somewhat 'leaky' shield which does not exhibit full suppression of electric field feedthrough. This design, though appearing suboptimal at the first glance, has the advantage of providing less effective area for the formation of eddy currents and thus less degradation of the coil's Q factor than a less leaky shield.
This trade-off was first explored in FE simulations which were then validated experimentally with the construction of the copper-coil shield.
The experimental feedthrough (\(\xi\)-value) is somewhat lower than theoretically expected from the simulation. Such small deviations are to be expected due to model simplifications such as the assumption of insulating boundary conditions on some faces where displacement currents still may flow and small deviations between true and modeled geometry. Overall, the FE simulations and experimental results demonstrate strong agreement in both shielding factor and Q factor for the copper-foil shield.
In our case, approximately 6 - 7\% of interference currents would still reach the coil. 
A non-segmented coil would still pick up a significant part of this signal especially at the 'hot' end of the coil, i.e. the one far from signal ground, but the combination of a shield with a segmented coil design can achieve very good overall suppression with a tolerable reduction in Q factor. 

With the basic concept confirmed in simulations and phantom experiments, we designed an improved shield based on flexible PCBs (FENCE).
This solution is preferable, due to its superior performance and mechanical reliability.
For FENCE, the loaded Q factor dropped by 9\% in the phantom experiments which is well tolerable. The Q factor drop is more pronounced in the unloaded configuration, as sample noise already contributes significantly to Q factor degradation when loaded. The overall drop in Q factor due to FENCE therefore also depends on the loading condition of the RF coil which can be observed when comparing the loaded Q factor values for phantom and in-vivo experiments.
Whether FENCE improves overall SNR therefore depends on measurement conditions. In environments with existing Faraday shielding and adequate subject grounding FENCE may decrease overall image quality as the SNR scales with the square root of Q for identical coil geometries \cite{Webb_2023}. 
For in-vivo measurements under minimal EMI conditions, FENCE could result in a theoretical 10\% SNR reduction due to the corresponding Q factor decrease.
However, when EMI becomes significant, FENCE can improve SNR despite lower RF coil Q factor. This effect was observed in phantom measurements where, even without active EMI introduction, ambient interference was measurable and mitigated by FENCE, resulting in SNR improvement.
This was then also confirmed in our in-vivo experiments, where SNR improved significantly  with and without subject grounding. 
It should be noted that there is also still room for improvement. Future work will focus on optimizing FENCE geometry to maximize shielding while minimizing eddy current losses. Additionally, the FENCE concept will be tested for shielding between gradient coils and RF coils as well as different coil geometries outside of simple solenoid coils.

FENCE reduces the need for bulky Faraday structures around patients as it effectively mitigates EMI. This could make low-field MRI more practical in space-constrained environments such as intensive care units. 
Future research could conduct experiments in fully shielded environments with controlled and realistic EMI sources representative of clinical settings, facilitating an implementation beyond research settings.
This would be a first step for the transition from research environments to clinical settings.
At approximately 100 EUR for 15 PCBs sufficient to equip multiple RF coils, FENCE provides a cost-effective solution for retrofitting existing coils to enhance EMI immunity.

Nevertheless, active noise cancellation may still remain important under harsh environmental conditions, especially in the presence of very strong magnetic EMI sources such as high power switched mode power supplies as evidenced by the in-vivo experiments.
The post-processing nature of these methods makes them rather flexible. Depending on environmental EMI conditions, they can be applied as needed, allowing direct observation of image quality improvements while preserving the original measurement data. By combining passive approaches such as subject grounding or FENCE with active EMI sensing, a complementary system can be achieved that is robust to both inductively and capacitively coupled EMI.
When applying active noise cancellation such as with EDITER, our results suggest that it is important to include electrical pickup sensors such as electrodes, which is in agreement with the findings of Srinivas et al. \cite{srinivas_external_2022}, where best results could be obtained only when using an electrode attached to the body as a reference input.
Future work can investigate the performance of active noise cancellation in combination with the FENCE approach and observe the effect of different sensor types.

\section{Conclusions}\label{sec5}

FENCE in combination with capacitive RF coil segmentation provides an efficient EMI reduction method without complex electronics or large Faraday shields. Existing low-field RF coils can be retrofitted with FENCE to enhance EMI immunity at low cost.

\section*{Author contributions}

JP: hardware design, experiments, data analysis, original draft, writing - review and editing.
HS: simulations, initial prototyping and experiments, interpretation of the result, writing - review and editing.
MU: supervision, project administration, funding acquisition, interpretation of the result, writing - review and editing.
All authors approved the final version of the manuscript.

\section*{Acknowledgments}
Thank you to the Open Source Imaging Initiative e.V. for providing design files for the various MRI components, and Lukas Winter and Martin Häuer for discussions regarding open-source low-field MRI scanner assembly. Thank you to the members of the OSII Matrix\footnote{\url{https://matrix.to/\#/\#osii:matrix.org}} community for valuable discussions regarding low-field MRI.
We also would like to thank the volunteer who participated in this study. 
Thank you to Umberto Zanovello for providing the design files of the head coil prior to its public release. 

\section*{Data Availability Statement}

The design files and code used for analysis of the experimental data used in this study can be found at \url{https://gitlab.tugraz.at/ibi/mrirecon/papers/fence}
The experimental data can be found at \url{https://repository.tugraz.at/records/j111b-6q129}.
In case you want to obtain your own FENCE PCB the ongoing development can be followed here: \url{https://gitlab.tugraz.at/ibi/mrirecon/hardware/fence}

\bibliography{shieldingpaper}

\begin{thebibliography}{10}

\bibitem{Zhao_2024}
Y.~Zhao, Y.~Ding, V.~Lau, C.~Man, S.~Su, L.~Xiao, A.~T.~L. Leong, and E.~X. Wu, ``Whole-body magnetic resonance imaging at 0.05 tesla,'' {\em Science}, vol.~384, May 2024.

\bibitem{Cooley_2020}
C.~Z. Cooley, P.~C. McDaniel, J.~P. Stockmann, S.~A. Srinivas, S.~F. Cauley, M.~Śliwiak, C.~R. Sappo, C.~F. Vaughn, B.~Guerin, M.~S. Rosen, M.~H. Lev, and L.~L. Wald, ``A portable scanner for magnetic resonance imaging of the brain,'' {\em Nature Biomedical Engineering}, vol.~5, pp.~229--239, Nov. 2020.

\bibitem{Yuen_2022}
M.~M. Yuen, A.~M. Prabhat, M.~H. Mazurek, I.~R. Chavva, A.~Crawford, B.~A. Cahn, R.~Beekman, J.~A. Kim, K.~T. Gobeske, N.~H. Petersen, G.~J. Falcone, E.~J. Gilmore, D.~Y. Hwang, A.~S. Jasne, H.~Amin, R.~Sharma, C.~Matouk, A.~Ward, J.~Schindler, L.~Sansing, A.~de~Havenon, A.~Aydin, C.~Wira, G.~Sze, M.~S. Rosen, W.~T. Kimberly, and K.~N. Sheth, ``Portable, low-field magnetic resonance imaging enables highly accessible and dynamic bedside evaluation of ischemic stroke,'' {\em Science Advances}, vol.~8, Apr. 2022.

\bibitem{Bian_2024}
W.~Bian, P.~Li, M.~Zheng, C.~Wang, A.~Li, Y.~Li, H.~Ni, and Z.~Zeng, ``A review of electromagnetic elimination methods for low-field portable {MRI} scanner,'' in {\em 2024 5th International Conference on Machine Learning and Computer Application (ICMLCA)}, pp.~614--618, IEEE, 2024.

\bibitem{srinivas_external_2022}
S.~A. Srinivas, S.~F. Cauley, J.~P. Stockmann, C.~R. Sappo, C.~E. Vaughn, L.~L. Wald, W.~A. Grissom, and C.~Z. Cooley, ``External {Dynamic} {InTerference} {Estimation} and {Removal} ({EDITER}) for low field {MRI},'' {\em Magnetic Resonance in Medicine}, vol.~87, no.~2, pp.~614--628, 2022.

\bibitem{liu_low-cost_2021}
Y.~Liu, A.~T.~L. Leong, Y.~Zhao, L.~Xiao, H.~K.~F. Mak, A.~C.~O. Tsang, G.~K.~K. Lau, G.~K.~K. Leung, and E.~X. Wu, ``A low-cost and shielding-free ultra-low-field brain {MRI} scanner,'' {\em Nature Communications}, vol.~12, no.~1, p.~7238, 2021.

\bibitem{zhao_electromagnetic_2024}
Y.~Zhao, L.~Xiao, Y.~Liu, A.~T. Leong, and E.~X. Wu, ``Electromagnetic interference elimination via active sensing and deep learning prediction for radiofrequency shielding‐free {MRI},'' {\em NMR in Biomedicine}, vol.~37, no.~7, p.~e4956., 2024.

\bibitem{zhao_robust_2024}
Y.~Zhao, L.~Xiao, J.~Hu, and E.~X. Wu, ``Robust {EMI} elimination for {RF} shielding‐free {MRI} through deep learning direct {MR} signal prediction,'' {\em Magnetic Resonance in Medicine}, vol.~92, no.~1, pp.~112--127, 2024.

\bibitem{Lu_2025}
R.~Lu, Z.~Wu, G.~Zhang, X.~Hu, Y.~Liu, X.~Jiang, Z.~Ni, and H.~Yi, ``Electromagnetic interference rejection strategy for 50-mt portable unshielded whole-body magnetic resonance imaging with a convolutional neural network incorporating attention mechanism,'' {\em IEEE Transactions on Instrumentation and Measurement}, vol.~74, pp.~1--8, 2025.

\bibitem{Negnevitsky_2023}
V.~Negnevitsky, Y.~Vives-Gilabert, J.~M. Algarín, L.~Craven-Brightman, R.~Pellicer-Guridi, T.~O’Reilly, J.~P. Stockmann, A.~Webb, J.~Alonso, and B.~Menküc, ``{MaRCoS}, an open-source electronic control system for low-field {MRI},'' {\em Journal of Magnetic Resonance}, vol.~350, p.~107424, 2023.

\bibitem{guallartnaval_2025}
T.~Guallart‐Naval, J.~M. Algarín, and J.~Alonso, ``Electromagnetic noise characterization and suppression in low‐field mri systems,'' {\em Magnetic Resonance in Medicine}, Jan. 2026.

\bibitem{Yang_2022}
L.~Yang, W.~He, Y.~He, J.~Wu, S.~Shen, and Z.~Xu, ``Active {EMI} suppression system for a 50 {mT} unshielded portable {MRI} scanner,'' {\em IEEE Transactions on Biomedical Engineering}, vol.~69, no.~11, pp.~3415--3426, 2022.

\bibitem{Qiao_2025}
S.~Qiao, Y.~Yu, T.~Lin, J.~Jiang, Y.~Liu, and X.~Li, ``Emi cancellation for shielding-free ultra-low-field mri,'' {\em IEEE Transactions on Biomedical Engineering}, pp.~1--11, 2025.

\bibitem{lena_subject_2025}
B.~Lena, B.~de~Vos, T.~Guallart‐Naval, J.~Parsa, P.~Garcia~Cristobal, R.~van~den Broek, C.~Najac, J.~Alonso, and A.~Webb, ``Subject grounding to reduce electromagnetic interference for {MRI} scanners operating in unshielded environments,'' {\em Magnetic Resonance in Medicine}, 2025.
\newblock Early View - Online Version of Record before inclusion in an issue.

\bibitem{Li_2017}
J.~Li, Z.~Nie, Y.~Liu, L.~Wang, and Y.~Hao, ``Evaluation of propagation characteristics using the human body as an antenna,'' {\em Sensors}, vol.~17, no.~12, p.~2878, 2017.

\bibitem{bronzino_bmtbook}
J.~H. Nagel, {\em Medical Instruments and Devices}, vol.~2 of {\em The Biomedical Engineering Handbook}.
\newblock CRC Press, Taylor \& Francis Group, fourth~ed., Apr. 2015.
\newblock Chapter 9 Biopotential Amplifiers - Section 9.2.1 Interferences.

\bibitem{antenna_theory_2005}
C.~A. Balanis, {\em Antenna Theory: Analysis Design}.
\newblock John Wiley \& Sons, Inc., Hoboken, New Jersey, third edition~ed., 2005.
\newblock Chapter 2 Fundamental Parameters of Antennas - Section 2.2.4 Field Regions.

\bibitem{Mispelter_2006}
J.~Mispelter, M.~Lupu, and A.~Briguet, {\em {NMR} Probeheads for Biophysical and Biomedical Experiments: Theoretical Principles and Practical Guidelines}.
\newblock Imperial College Press London, first~ed., May 2006.
\newblock Chapter 10 Probe Evaluation and Debugging - Section 10.1.2.2 Evaluation of Q factor.

\bibitem{Gadian_1979}
D.~Gadian and F.~Robinson, ``Radiofrequency losses in {NMR} experiments on electrically conducting samples,'' {\em Journal of Magnetic Resonance (1969)}, vol.~34, pp.~449--455, May 1979.

\bibitem{Wu_2009}
C.~H. Wu, C.~V. Grant, G.~A. Cook, S.~H. Park, and S.~J. Opella, ``A strip-shield improves the efficiency of a solenoid coil in probes for high-field solid-state {NMR} of lossy biological samples,'' {\em Journal of Magnetic Resonance}, vol.~200, pp.~74--80, Sept. 2009.

\bibitem{Krahn_2008}
A.~Krahn, U.~Priller, L.~Emsley, and F.~Engelke, ``Resonator with reduced sample heating and increased homogeneity for solid-state nmr,'' {\em Journal of Magnetic Resonance}, vol.~191, pp.~78--92, Mar. 2008.

\bibitem{Park_2010}
B.~Park, T.~Neuberger, A.~G. Webb, D.~C. Bigler, and C.~M. Collins, ``Faraday shields within a solenoidal coil to reduce sample heating: Numerical comparison of designs and experimental verification,'' {\em Journal of Magnetic Resonance}, vol.~202, pp.~72--77, Jan. 2010.

\bibitem{conradi_rf_2019}
M.~S. Conradi and A.~P. Zens, ``{RF} shielding and eddy currents in {NMR} probes,'' {\em Journal of Magnetic Resonance}, vol.~305, pp.~180--184, 2019.

\bibitem{Webb_2023}
A.~Webb and T.~O’Reilly, ``Tackling {SNR} at low-field: a review of hardware approaches for point-of-care systems,'' {\em Magnetic Resonance Materials in Physics, Biology and Medicine}, vol.~36, no.~3, pp.~375--393, 2023.

\bibitem{COMSOL_documentation}
COMSOL Multiphysics, {\em AC/DC Module User's Guide, Transition Boundary Condition}, v. 6.2~ed., 2023.
\newblock Available at: \url{https://doc.comsol.com/6.2/doc/com.comsol.help.woptics/woptics_ug_optics.6.26.html}.

\bibitem{Yang2014}
P.~Yang, F.~Tian, and Y.~Ohki, ``Dielectric properties of poly(ethylene terephthalate) and poly(ethylene 2,6-naphthalate),'' {\em IEEE Transactions on Dielectrics and Electrical Insulation}, vol.~21, pp.~2310 -- 2317, 2014.

\bibitem{Konieczna2010}
M.~Konieczna, E.~Markiewicz, and J.~Jurga, ``Dielectric properties of polyethylene terephthalate/polyphenylene sulfide/barium titanate nanocomposite for application in electronic industry,'' {\em Polymer Engineering \& Science}, vol.~50, pp.~1613--1619, 2010.

\bibitem{osii_handcoil}
C.~Engler, R.~Montag, T.~O'Reilly, S.~Schachel, and L.~Winter, ``{Solenoid d150 N20}.'' GitLab, 2025.
\newblock https://gitlab.com/osii/rf-system/rf-coils/solenoid-d150-n20, commit: 75a442d1c09e105c508fce087247533c2d21d67c.

\bibitem{osii_headcoil}
U.~Zanovello, ``{A4IM RF Head Coil}.'' GitLab, 2025.
\newblock https://gitlab.com/osii/rf-system/rf-coils/a4im-rf-head-coil, commit: fd0f42816851079b292181cb2a0368800c05d4b2.

\bibitem{tao_segmented_2023}
Y.~Tao, R.~Lemdiasov, A.~Venkatasubramanian, and M.~Wong, ``{Segmented Coil Design Powering the Next Generation of High-efficiency Robust Micro-implants},'' in {\em Smart Grids Technology and Applications} (L.~Mihet-Popa, ed.), ch.~6, London: IntechOpen, 2022.

\bibitem{osii_one}
{Open Source Imaging Initiative}, ``{OSI² ONE: Open Source low-field MRI scanner}.'' Gitlab, 2023.
\newblock https://gitlab.com/osii/mri-scanners/osii-one.

\bibitem{ocra}
T.~Witzel, ``{OCRA MRI}.'' Website.
\newblock https://openmri.github.io/ocra/, Accessed 20-11-2025.

\bibitem{osii_console}
J.~Frintz, M.~Häuer, C.~Höhn, M.~Tobias, R.~Montag, D.~Schote, F.~Seifert, B.~Silemek, and L.~Winter, ``{RedPitConsole}.'' GitLab, 2023.
\newblock https://gitlab.com/osii/console/redpitconsole, commit: dccdfff9369428e95587b24be15305c6d5681d8c.

\bibitem{osii_rfpa}
D.~de~Gans, ``{1kW peak RFPA}.'' GitLab, 2022.
\newblock https://gitlab.com/osii/rf-system/rf-power-amplifier/1kw-peak-rfpa, commit: 12db0c05518f871c5a809aac88f7eb013608d5b2.

\bibitem{osii_gpa}
D.~de~Gans, ``{TU Delft GA2}.'' GitLab, 2022.
\newblock https://gitlab.com/osii/gradient-system/gradient-power-amplifier/tu-delft-ga2, commit: e49ae763f2266c0a2a35dda1ec5940065402f7a4.

\bibitem{osii_magnet}
M.~Häuer, T.~O'Reilly, S.~Schachel, W.~Teeuwisse, and L.~Winter, ``{30cm Halbach Magnet}.'' GitLab, 2024.
\newblock https://gitlab.com/osii/magnet/30cm-halbach-magnet, commit: 1b5b726184550a9ac52e231497d3e05f383ec547.

\bibitem{a4im_txrx}
M.~Prier, ``{A4IM OSI² TR-Switch}.'' Website, 10 2024.
\newblock https://zeugmatographix.org/ocra/2024/10/11/a4im-osi

\bibitem{Algar_n_2024}
J.~M. Algarín, T.~Guallart-Naval, J.~Borreguero, F.~Galve, and J.~Alonso, ``{MaRGE}: {A} graphical environment for {MaRCoS},'' {\em Journal of Magnetic Resonance}, vol.~361, p.~107662, 2024.

\bibitem{Gudbjartsson_1995}
H.~Gudbjartsson and S.~Patz, ``The rician distribution of noisy {MRI} data,'' {\em Magnetic Resonance in Medicine}, vol.~34, no.~6, pp.~910--914, 1995.

\bibitem{Han_2017}
H.~Han, R.~Moritz, E.~Oberacker, H.~Waiczies, T.~Niendorf, and L.~Winter, ``Open source {3D} multipurpose measurement system with submillimetre fidelity and first application in magnetic resonance,'' {\em Scientific Reports}, vol.~7, no.~1, p.~13452, 2017.

\end{thebibliography}
\bibliographystyle{ieeetr}

\newpage

\appendix

\section{In-Vivo Measurement Setup}

\begin{figure}[h]
    \centering
    \includegraphics[width=1.0\linewidth]{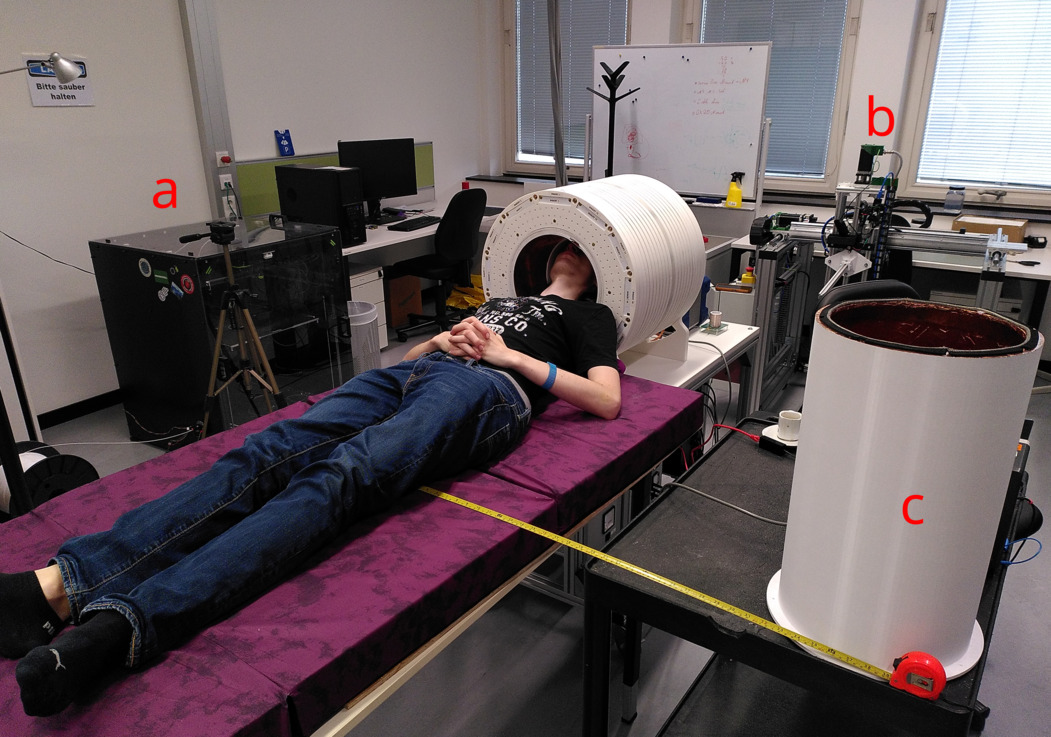}
    \caption{Measurement setup for the in-vivo measurements. Different EMI sources are highlighted in red. (a) shows the 3D printer (b) the field mapping robot and (c) the copper foil cylinder for EMI coupling. }
    \label{fig:measurement_setup}
\end{figure}

\end{document}